\documentclass[useAMS,usenatbib]{mn2e}
\usepackage[pdftex]{graphicx,color}
\usepackage[latin1]{inputenc}
\usepackage{graphics}
\usepackage{amsfonts}
\usepackage{amsmath}
\usepackage{multicol}
\usepackage{layout}
\usepackage{amssymb}
\usepackage[a4paper,colorlinks=true,pdfstartview=FitV,
linkcolor=red,citecolor=blue,urlcolor=magenta]{hyperref}
\title[Halo model at $k \gg 1$ Mpc$^{-1}$]
  {Halo model description of the non-linear dark matter power spectrum at $k \gg 1$ Mpc$^{-1}$}
\author[Giocoli et al. 2010]{\parbox{\textwidth}{Carlo Giocoli$^{1}$\thanks{E-mail:
\href{mailto:cgiocoli@ita.uni-heidelberg.de}{cgiocoli@ita.uni-heidelberg.de}},
    Matthias Bartelmann$^1$, Ravi K.~Sheth$^2$,
    Marcello Cacciato$^{3,4}$\thanks{Minerva Fellow}}\\ \\
  $^1$Zentrum f\"ur Astronomie, ITA, Universit\"at Heidelberg,
  Albert-Ueberle-Str. 2, 69120 Heidelberg, Germany \\
  $^2$Center for Particle Cosmology, University of Pennsylvania,
  209 S. 33rd Street Philadelphia, PA 19104-6396 \\
  $^3$Racah Institute of Physics, The Hebrew University Jerusalem,
  91904 Israel \\
  $^4$Max-Planck-Institute for  Astronomy, K\"onigstuhl  17, D-69117
  Heidelberg, Germany}

\begin{document}
\date{}
\maketitle
\label{firstpage}
\pagerange{\pageref{firstpage}--\pageref{lastpage}} \pubyear{2010}

\begin{abstract}
  Accurate knowledge  of the non-linear dark-matter  power spectrum is
  important  for  understanding   the  large-scale  structure  of  the
  Universe, the statistics of  dark-matter haloes and their evolution,
  and cosmological  gravitational lensing.  We  analytically model the
  dark-matter  power  spectrum   and  its  cross-power  spectrum  with
  dark-matter  haloes.  Our  model extends  the  halo-model formalism,
  including   realistic  substructure  population   within  individual
  dark-matter haloes and the scatter of the concentration parameter at
  fixed  halo   mass.   We   consider  three  prescriptions   for  the
  mass-concentration   relation   and   two   for   the   substructure
  distribution in  dark-matter haloes. We show that  this extension of
  the halo  model mainly  increases the predicted  power on  the small
  scales,  and  is  crucial  for  proper  modeling  the  cosmological
  weak-lensing signal due to  low-mass haloes.  Our extended formalism
  shows how the halo model approach can be improved in accuracy as one
  increases  the  number  of  ingredients  that  are  calibrated  from
  $n$-body simulations.
\end{abstract}
\begin{keywords}
  galaxies:  halos  -  cosmology:  theory  - dark  matter  -  methods:
  analytical - gravitational lensing: weak
\end{keywords}

\section{Introduction}

Accurate   studies  of  large-scale   cosmic  structures   and  galaxy
clustering became  possible with the  advent of large  galaxy redshift
surveys \citep{guzzo05,meneux09}. Different analyses have been carried
out   depending   on   galaxy   luminosity,  color   and   morphology
\citep{norberg02,zehavi05,padmanabhan07}.   According to  the standard
scenario  of structure  formation, galaxies  with  dissimilar features
reside  in different dark  matter haloes  and have  experienced varied
formation histories.

Galaxies are  believed to form  and reside in dark-matter  haloes that
extend much beyond their observable radii. Some of them are located at
halo centers while others  orbit around it, constituting the satellite
population. Dark-matter haloes  form by gravitational instability from
dark-matter    density    fluctuations   \citep{bond91,lacey93}    and
subsequently merge  to form increasingly  large haloes as  cosmic time
proceeds. Gas  follows the dark-matter density  perturbations. Once it
reaches sufficiently high densities, dissipative processes, shocks and
cooling allow stars to form from this gas \citep{white78,kauffmann99}. 
While the main lines of this scenario are widely accepted, its details 
are still poorly understood.

The clustering  strength of  a given galaxy  population is  related to
that of the dark matter halos which host the galaxies.  It is possible
to provide  an accurate description  of clustering in  the small-scale
non-linear  regime even  if one  has no  knowledge of  how  the haloes
themselves are clustered  \citep{sheth97b,smith03}.  The halo-model of
matter                                                       clustering
\citep{scherrer91,peacock00,seljak00,scoccimarro01,cooray02},     which
has been the subject of much recent interest, allows a parametrization
of clustering even on large scales.

To date,  almost all  analytic work based  on the  halo-model approach
assumes  that haloes  are spherically  symmetric and  that  the matter
density  distribution around  each  halo centre  is smooth.   However,
numerical  simulations  of  hierarchical  clustering have  shown  that
haloes         are        neither         spherically        symmetric
\citep{jing02,allgood06,hayashi07}              nor             smooth
\citep{moore98,springel01b,gao04,delucia04,tormen04,giocoli08b}.
About $10\%$  of the mass  in cluster-sized haloes is  associated with
subclumps.   A   halo  model  which  includes  the   effects  of  halo
triaxiality   on  various  clustering   statistics  is   developed  in
\cite{smith05,smith06},  and formalism  to include  halo substructures
was developed by \cite{sheth03c}.
 
The main  purpose of this work  is to include recent  advances in our
understanding  of halo  substructures into  the halo  model approach.
This is  because an  accurate model of  substructures is  a necessary
first  step to modeling  the small  scale weak  gravitational lensing
convergence and  shear signals \citep{bartelmann01,  hagan05} and for
the   substructure   contribution   to  the   gravitational   flexion
\citep{bacon06}.

The present  paper is organized as  follows.  In Sect.~\ref{themodel},
we describe  the ingredients of our  extension of the  halo model: the
properties  of  subhaloes  and  the substructure  mass  function.   In
Sect.~\ref{nonlinearps},  we show  how to  incorporate these  into the
halo  model.  Cross-correlations between  haloes and  mass as  well as
clumps   and  mass   are  studied   in  Sect.~\ref{crosscorrelations}.
Sect.~\ref{sandc} summarizes our methods and conclusions.

\section{The model}
\label{themodel}
We assume  a flat $\mathrm{\Lambda}$CDM  cosmological model consistent
with  a   combined  analysis  of  the   2dFGRS  \citep{colless01}  and
first-year  WMAP  data   \citep{spergel03}.   It  has  the  parameters
($\Omega_m$, $\Omega_b$, $h$,  $n$, $\sigma_8$)=(0.25, 0.045, 0.73, 1,
0.9).   This  is the  model  adopted  by  \citet{springel05b} for  the
Millennium Simulation (hereafter MS), against which we test our model.
Our   linear   power   spectrum   uses  the   transfer   function   by
\citet{eisenstein98a}, which takes baryonic acoustic oscillations into
account, in agreement with \textsc{CMBFAST} \citep{seljak96} to $1\%$.

\subsection{Halos}

\subsubsection{Mass function}

For modeling the non-linear dark-matter power spectrum, we require the
redshift  evolution  of  the   number  density  of  collapsed  haloes,
$n(M,z)$.   According  to  the  spherical collapse  model,  a  density
fluctuation  collapses when  and  where the  linearly evolved  density
field  smoothed  with a  top-hat  filter  exceeds  the critical  value
$\delta_c(z)$  at that  redshift. The  distribution of  collapsed peak
heights  is   Gaussian  when  expressed  in  terms   of  the  variable
$\mu(M,z)=\delta_c(z)/\sqrt{S(M)}$, where $S(M)$  is the variance of a
halo of mass $M$, if the collapse is not influenced by the surrounding
gravitational field \citep{press74,  bond91, lacey93}.  Following this
idea, one can write the halo number density in the form
\begin{equation}
 \nu f(\nu) = \dfrac{M^2}{\bar{\rho}} n(M,z)
 \dfrac{\mathrm{d}\log(M)}{\mathrm{d}\log(\nu)}\,,
\end{equation}
where  $\nu=\mu^2$  ($f(\mu)\mathrm{d}\mu$  is  Gaussian for  a  fixed
threshold $\delta_c$), and $\bar{\rho}$  is the mean matter density of
the Universe.

However,  the dark-matter  halo  mass functions  measured in  $N$-body
simulations  are  far  from  Gaussian.   The  \citet{press74}  formula
over-predicts the abundance of low-mass haloes and under-predicts that
of high-mass  systems \citep{sheth99b}.  This is  because the collapse
of  the haloes is  affected by  the gravitational  tidal field  of the
surrounding matter \citep{sheth01b}.   Different functional forms have
been      proposed     to      fit     the      simulation     results
\citep{sheth99b,jenkins01}.  We use the \citet{sheth99b} mass function
because  it is well  tested against  numerical simulations  and easily
parametrised  in  terms of  the  variable  $\nu$  which we  shall  use
throughout.

\subsubsection{Bias}

Since the  halo model assumes  that all matter  in the universe  is in
collapsed  systems, we  need  to  model the  ratio  between the  power
spectra of  dark-matter haloes and  of the dark matter,  i.e.~the bias
parameter, to compute the dark-matter power spectrum. To be consistent
with our choice  of the mass function, we also  use the bias parameter
$b(\nu)$ derived  by \citet{sheth99b}. When expressed in  terms of the
peak height $\nu$,  the bias parameter is independent  of redshift. We
recall  that  both  the  mass  function  and  the  bias  parameter  by
\citet{sheth99b} reproduce  the original equations  by \citet{press74}
and \citet{mo96}  when specialized to isolated  spherical collapse. At
fixed redshift, the  bias is an increasing function  of the halo mass.
More massive  haloes (forming later and having  a lower concentration)
are more biased  compared to less massive haloes  (forming earlier and
with  a  higher concentration).  This  has  been  confirmed by  recent
$N$-Body simulations.

Different  improvements   of  the  bias  factor   have  been  proposed
\citep{jing98, governato99b,  sheth01b, mandelbaum05}. Recent analyses
have shown interesting features and  unexpected results in view of the
standard excursion-set  formalism \citep{sheth04b}, also  confirmed by
\citet{gao05a, faltenbacher10}.  At  masses exceeding $M_*$ (i.e.~peak
heights  $\nu>1$),  haloes  in  the  same  mass  bin  but  with  lower
concentration (and thus lower formation redshift) are more biased than
those with high concentrations, with a scatter of $\approx20\%$ around
the  mean. The  opposite holds  for masses  less than  $M*$ ($\nu<1$).
There is  currently no accurate  analytic model which allows  a simple
parametrization of the  `assembly bias' effect, so we  have decided to
continue using a  bias parameter that depends on  halo mass alone. Our
algorithm can be easily extended by a new model for the bias factor if
needed.

\subsubsection{Density profiles}

\begin{figure}
\centering
\includegraphics[width=\hsize]{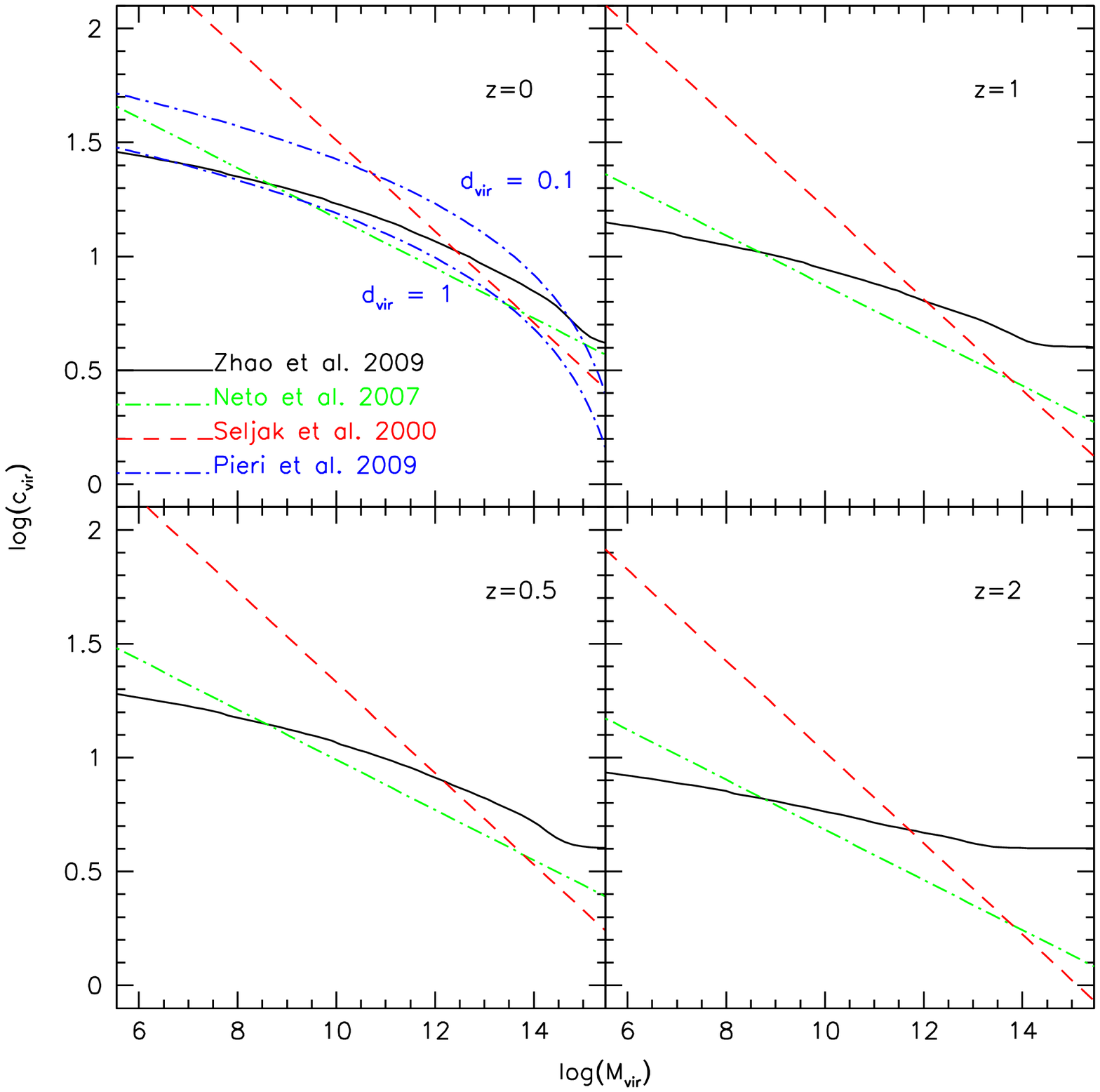}
\caption{Mass-concentration  relations of  dark-matter haloes  at four
  different redshifts.  The solid,  dash-dotted and dashed curves show
  the models by \citet{zhao09}, \citet{neto07} and \citet{seljak00} at
  redshifts $z$=0, 0.5, 1 and  2, respectively. In the top-left panel,
  we  also  show  the  mass-concentration relation  for  substructures
  estimated  by \citet{pieri09}  fitting the  results of  the Aquarius
  simulation for clumps located at the virial radius ($d_{vir}=1$) and
  at $0.1  R_{vir}$ from the host-halo  centre ($d_{vir}=0.1$). Notice
  that  this fit has  been rescaled  to the  definition of  the virial
  radius adopted in this work. \label{mcfig}}
\end{figure}

Various  different definitions for  the virial  overdensity $\Delta_h$
have  been used  in  the literature.   Some authors  \citep{jenkins01,
  gao04}  use  $\Delta_{200c}$,   corresponding  to  $200$  times  the
\textit{critical} density, or  $\Delta_{200b}$, defined as $200$ times
the    \textit{background}    density   \citep{diemand07b}.     Others
\citep{sheth99b,     sheth01b,      bullock01a,     jing02}     choose
$\Delta_h=\Delta_{vir}$  according  to  the spherical  collapse  model
\citep{peebles80,    kitayama96,    bryan98}.     Here,    we    adopt
$\Delta_h=\Delta_{vir}$  and  use the  virial  overdensities given  by
\citet{eke96}, which are related to the virial mass $M_{vir}$ by
\begin{equation}
 M_{vir} = \frac{4 \pi}{3} R_{vir}^3 \frac{\Delta_{vir}}{\Omega_m(z)}
 \Omega_0 \rho_c \,.
\end{equation}

Cosmological  $N$-body simulations \citep{navarro96,  power03, neto07,
  gao08} have  shown that the  dark-matter density in  isolated haloes
has a universal radial profile whose slope steepens from $-1$ near the
centre to $-3$ towards the virial radius,
\begin{equation}
 \rho(r|M_{vir}) = \dfrac{\rho_s}{r/r_s\left( 1 + r/r_s\right)^2}\,,
\end{equation}
where  $r_s$ is the  scale radius  of the  halo. Its  concentration is
defined as  $c_{vir} = R_{vir}/r_s$.  The amplitude $\rho_s$  sets the
dark-matter density at the scale radius,
\begin{equation}
 \rho_s = \dfrac{M_{vir}}{4 \pi r_s^3} \left[ \ln(1+c_{vir}) -
 \dfrac{c_{vir}}{1+c_{vir}}\right]^{-1}\;.
\end{equation}
Note that $c_{vir}$,  and thus also  $\rho_s$, depend  on halo mass; 
we expand on this in the next subsection.

We  shall need  in  the  following an  expression  for the  normalized
Fourier  transform of this  density profile,  truncated at  the virial
radius,
\begin{equation}
 u(k|M) = \int_0^{R_{vir}} \frac{4\pi r^2}{M} \frac{\sin kr}{kr}
 \rho(r|M)\,\mathrm{d}r\;.
\end{equation}
For the NFW profile, this integral has a convenient analytic
expression \citep[see][]{scoccimarro01}. 

\subsubsection{Mass-concentration relations}

Motivated by numerical simulations and by previous halo-model studies,
we consider three prescriptions for the mass-concentration relation to
study uncertainties and differences  in the recovery of the non-linear
dark-matter power spectrum by means of the halo model \citep{seljak00,
  cooray02}. These are:

\begin{enumerate}

\item \textbf{C1} \citep{neto07}:
\begin{equation}
 c^N_{vir}(z) = \frac{c_{14}}{(1+z)} \left[ \frac{M}{10^{14}}\right]^{-0.11}\;.
\label{netoeq}
\end{equation}
This relation  was obtained by \citet{neto07} by  fitting the complete
sample   of   haloes   at   $z=0$   in   the   Millennium   Simulation
\citep{springel05b}.    The   value   of   the  concentration   of   a
$10^{14}M_{\odot}/h$ halo estimated  by \citet{neto07} for an enclosed
mean density of $200$ times  the critical density has been rescaled to
the virial overdensity. To  rescale the normalization of the relation,
we  use the  relation  between $c_{200}$  and  $c_{vir}$ presented  in
\citet{maccio08} at $z=0$.

\item \textbf{C2} \citep{seljak00}:
\begin{equation}
 c^S_{vir}(z) = \frac{9}{(1+z)} \left[ \frac{M}{M_{*0}} \right]^{-0.2}\;.
\label{seljakeq}
\end{equation} 
This steeper model was used by  \citet{seljak00} in  order to
follow  the   \citet{peacock96}  fit   for  the  standard   CDM  model
adopted. The  mass $M_{*0}$ represents  the typical collapsed  mass at
$z=0$ such that $\delta_c(0)=\sigma(M_{*0})$.  Here, $\sigma(M)$ is the 
rms fluctuation in the  linear density-fluctuation field when smoothed 
with a top-hat filter  which  contains mass  $M$.  In this  case,  we  
set the concentration  of an  $M_*$  halo to  $9$  in order  to reproduce  
the non-linear   power-spectrum   of    the   Millennium   Simulation   at
$z=0$.  Models C1  and C2  incorporate the  redshift evolution  of the
mass-concentration relation proposed by \citet{bullock01a}.

\item \textbf{C3} \citep{zhao09}:
\begin{equation}
 c^Z_{vir}(z) = 4 \left[ 1 + \left( \frac{t(z)}{3.75 \, t_{0.04}}
 \right)^{8.4} \right]^{1/8}\;.
\label{zhaoeq}
\end{equation}
This  model was proposed  by  \citet{zhao09}   studying  the
mass-accretion history of dark-matter haloes in different cosmological
models.  They  developed a  model for the  mass accretion  history and
related the concentration of an  isolated halo to the time or redshift
$z$  when  it  had  assembled  $4\%$  of  its  final  mass.   This  is
qualitatively  motivated by the  expectation that  the structure  of a
halo should  depend on its  formation history. The correlation  of the
halo concentration with the time $t_{0.04}$ is purely empirical. For a
detailed explanation of  how to implement model C3,  see Appendix~A of
\citet{zhao09}.   So far,  the model  lacks a detailed physical 
justification.  However, it expresses the notion that dark-matter 
haloes form first as consequence of a violent relaxation process 
\citep{white96} which sets the  value  of  the  concentration  near $4$.
Subsequent accretion adds material primarily to the outer shells.   
A minimum value of $\sim 4$ means that the mass-concentration relation 
is expected to flatten at high redshifts, where haloes have had only 
had time to undergo the violent relaxation process.

This flattening  of the  mass-concentration relation at  high redshift
was  first discussed by  \citet{zhao03b}.  Based  on a  combination of
$N$-body simulations  with different resolutions, they  found that the
mass  dependence  of   halo  concentrations  weakens  with  increasing
redshift.   Later, \citet{gao08}  and \citet{duffy08}  confirmed these
results  studying  haloes at  different  redshifts  in the  Millennium
Simulation  and  with  a   different  set  of  numerical  simulations,
respectively.  They  proposed  different  power-law  fits  at  various
redshifts, but these are less complete than model C3.

\end{enumerate}

\begin{figure*}
\centering
\includegraphics[width=\hsize]{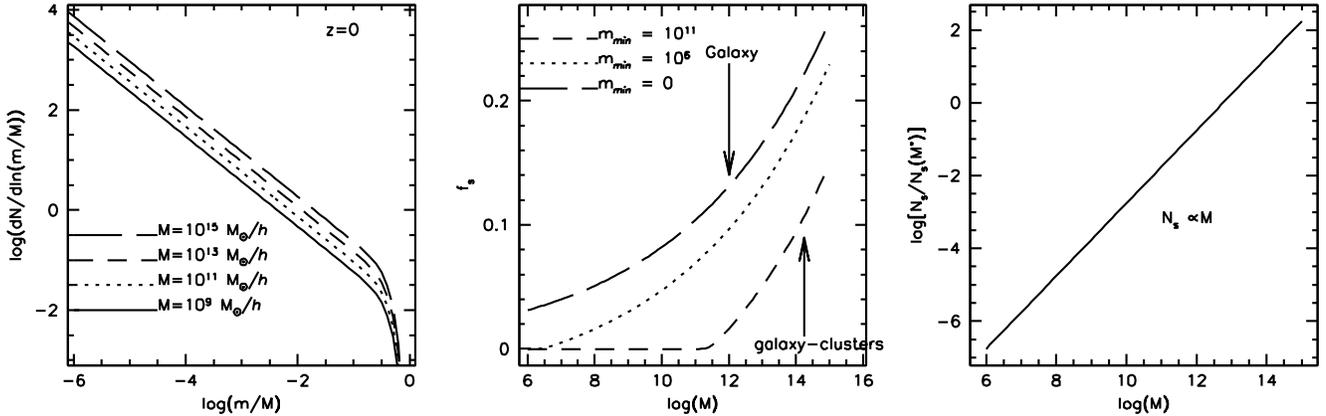}
\caption{Left:  Substructure mass  function  at the  present time  for
  different host-halo masses.  Middle: Dependence of substructure mass
  fraction  $f_s$ on host-halo  mass.  Curves  show $f_s$  in subhalos
  above the minimum mass, as  labeled.  Right: Dependence of number of
  substructures on host halo mass, expressed in units of the number in
  an $M_*$-halo. The solid line has a slope of unity. \label{subf}}
\end{figure*}

In   Fig.~\ref{mcfig},  we   show  the   mass-concentration  relations
described  above at four  different redshifts,  $z=0$, $0.5$,  $1$ and
$2$.  While  models C1  and C2 preserve  their slopes as  the redshift
increases, model  C3 flattens  towards higher redshift.   Also, notice
that in model  C3, no haloes have a concentration  $<4$, the value set
by the violent relaxation process.

\subsubsection{Concentration distributions}

We shall adopt two ways of assigning concentrations to haloes of fixed
mass, a deterministic model D1 and a stochastic model D2.

\begin{enumerate}

\item  \textbf{D1}: In  common  applications of  the  halo model,  any
  scatter  in  the  concentration  parameter  at fixed  halo  mass  is
  neglected. The conditional distribution  $p(c|M)$ is thus assumed to
  be a delta distribution,
\begin{equation}
 p(c|M)=\delta\left[c-\bar c(M, z)\right]\;,
\label{eq:D1}
\end{equation} 
where $\bar c(M, z)$ is given by any of the mass-concentration
relations C1, C2 or C3, see Eqs.~(\ref{netoeq}, \ref{seljakeq},
\ref{zhaoeq}).

\item \textbf{D2}: However, numerical simulations have shown that at a
  fixed halo mass, different assembly histories yield different values
  of    the    concentration    \citep{navarro96,jing00,wechsler02}. 
  For example, \citet{zhao03a,zhao03b} proposed that the concentration 
  could  be related  to  the  phase of fast accretion of the main halo
  progenitor. In  general, haloes formed  at higher redshift  are more
  concentrated  than  haloes formed  more  recently. The  distribution
  around the  mean value $\bar{c}$  found in numerical  simulations is
  well described by a log-normal form,
\begin{equation}
 p(c|M) = \frac{1}{\sqrt{2 \pi \sigma_{\ln c}^2}}
 \exp\left\{- \frac{\left[\ln(c/\bar{c}(M,z)) \right]^2}{2 \sigma_{\ln c}^2}\right\}\,,
\label{eq:D2}
\end{equation}
with variance $\sigma_{\ln c}$.  Values found in numerical simulations
are  $0.1\lesssim\sigma_{\ln c}\lesssim0.25$ 
 \citep{jing00, sheth04b, dolag04, neto07}.

\end{enumerate}

Given a concentration distribution $p(c|M)$, we define the
marginalised density profile in Fourier space
\begin{equation}
 \bar{u}(k|M) = \int p(c|M) u(k|c(M)) \mathrm{d}c\;,
\end{equation}
where  $\bar{c}$ is  again given  by any  of  the Eqs.~(\ref{netoeq}),
(\ref{seljakeq})  or  (\ref{zhaoeq}).   Note  that  $\bar{u}(k|M)$  is
different from $u(k|\bar{c}(M))$.

\subsection{Subhaloes}

\subsubsection{Abundance and mass function}

Increasing  the  mass  resolution  in  numerical  simulations,  recent
studies have found that cores  of accreted satellite haloes survive in
host  haloes  as  orbiting  substructures  \citep{tormen97a,  moore99,
  ghigna00,  springel01b}. Different  studies have  been  performed in
recent  years,  modeling  the  substructure  mass  function  and  its
dependence on the mass and the redshift of the host halo \citep{gao04,
  delucia04, giocoli08b, angulo09, giocoli10}.  Most of them find that
the number of substructures of mass $m$, at fixed mass fraction $m/M$,
depends  on  the host-halo  mass:  more  massive  haloes contain  more
subhaloes  than less  massive  haloes.  This  has  been translated  by
\citet{gao04} into the statement  that the number of substructures per
host-halo   mass   is    universal,   recently   also   confirmed   by
\citet{giocoli10}   using  different   post-processing  of   the  same
simulation. More  massive haloes  assemble their mass  quite recently,
thus the time spent by  their substructures under the influence of the
gravitational field of  the host halo is shorter  than in less massive
hosts.  \citet{giocoli10}  showed that this  also holds for  haloes of
similar mass,  but different  formation times. Systems  assembled more
recently (thus with a lower concentration) contain more substructures.

In the  model of \citet{giocoli10}, the  number of substructures
per halo mass can be written as
\begin{eqnarray}
 \frac{1}{M}   \frac{\mathrm{d}N(c(M),z)}{\mathrm{d}\ln   m}  &\equiv&
 \frac{\mathrm{d}N_M(c(M),z)}{\mathrm{d}\ln      m}      \label{subeq}
 \\  \nonumber   &=&  (1+z)^{1/2}  \frac{\bar{c}}{c}   A_M  m^{\alpha}
 \exp{\left[- \beta \left( \frac{m}{M} \right)^3\right]}\,,
\end{eqnarray}
where $A_M=9.33 \times 10^{-4}$  is a normalization factor and $\alpha
=-0.9$ and  $\beta=12.2715$ are the power-law slope  and the steepness
of the exponential cut-off.  The term $(1+z)^{1/2}$ takes the redshift
evolution  of   the  normalization   into  account.   It   depends  on
$c(M_{*z})/c(M_{*0})$,  while $\bar{c}/c$  reflects the  dependence of
the  normalization  on  the  concentration  at  fixed  host-halo  mass
\citep{giocoli10}.   Hereafter, subhalos within  the virial  region of
their   host  will   be  called   clumps.   In   the  left   panel  of
Fig.~\ref{subf}, we  show the substructure  mass function at  $z=0$ in
different host haloes.  In the  central panel we show the substructure
mass fraction in terms of the host halo mass,
\begin{equation}
 f_s = \int_{m_{min}}^M m
 \frac{\mathrm{d}N_M}{\mathrm{d}m}\mathrm{d}m\,,
\end{equation}
where  three  different  values  of  the minimum  mass  $m_{min}$  are
considered. In the right  panel, we show the halo-occupation statistic
of our model at $z=0$, normalized by the number of subhaloes in a halo
of mass $M_*$,
\begin{equation}
 \langle N_s\rangle = \int_0^M M \frac{\mathrm{d}N_M}{\mathrm{d}m}\mathrm{d}m\,,
\end{equation}
which  is  $\propto  M$   and has 
$\langle N_s(N_s-1)\rangle/\langle N_s\rangle^2=1$.

\subsubsection{Spatial distribution}

Next, we need a model  for the spatial distribution of subclumps around
the host-halo centre. We study two spatial density distributions:

\begin{enumerate}

\item \textbf{S1}:  The first assumes  that the clumps follow  the NFW
  dark matter distribution;

\item \textbf{S2}:  the second is motivated  by numerical simulations,
  assuming that the subclump distribution is less concentrated than the
  dark matter \citep{vandenbosch04, gao04}.

\end{enumerate}

\begin{figure}
\centering
\includegraphics[width=\hsize]{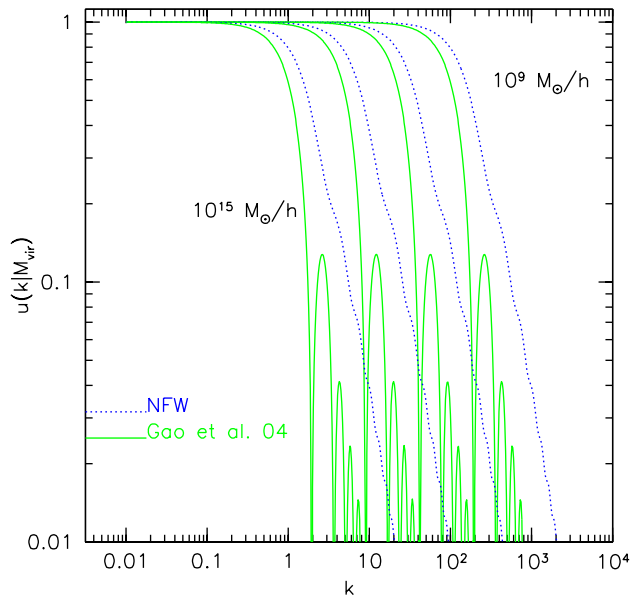}
\caption{\label{ukcf}  Fourier transform of  the density  profile. The
  dotted curves show the analytic Fourier transform of the NFW profile
  of different  host-halo masses: $10^{15}$,  $10^{13}$, $10^{11}$ and
  $10^9\,M_{\odot}/h$.   The solid curves  show the  numerical Fourier
  transform     of    the    \citet{gao04}     substructure    density
  distribution. The mass-concentration  relation by \citet{zhao09} was
  adopted here.}
\end{figure}

We shall later present results using  the two models S1 and S2 for the
substructure distribution  in the host  to illustrate by how  much the
power spectrum  changes accordingly. The  normalized Fourier transform
of the subclump distribution in a spherically symmetric system is
\begin{equation}
 U_s(k|c(M)) = \int_0^{R_{vir}} \frac{4 \pi r^2}{N_s} \frac{\sin k
 r}{kr} n_s(r,c(M)) \mathrm{d} r\,. \label{eqclumpdist}
\end{equation}
For  model S1,  $U_s(k|M)  =  u(k|M)$. For  model  S2, we  numerically
transform  the substructure  number-density  profile $n_s(r,M(c))$  of
\citet{gao04}. In  Fig.~\ref{ukcf}, we  show the Fourier  transform of
the  normalized density  distribution  for these  two profiles.   This
shows that  the \citet{gao04} fit drops  at much smaller  $k$ than the
NFW  profile, because  of the  different  trends of  the two  profiles
towards the host-halo centre.  We shall show in the next sections that
this affects the non-linear  power-spectrum at large $k$.  The ringing
in the Fourier  transform is due to the fact that  NFW is steeper than
\citet{gao04} both in the center and at the virial radius.

\subsubsection{Mass-concentration relation}

The  final  ingredient we  need  to  model  is the  mass-concentration
relation   for   the  substructures.   In   the   top-left  panel   of
Fig.~\ref{mcfig}, we  show the mass-concentration relation  fit to the
substructures  in  the   Aquarius  simulation  \citep{springel08b}  by
\citet{pieri09},
\begin{equation}
 c_{sub}(m,d_{vir}) = d_{vir}^{- \alpha_R}
 \left[ c_1 \;m_{sub}^{-\alpha_{c_1}} + c_2 
 \;m_{sub}^{-\alpha_{c_2}}\right]\,. \label{pierieq}
\end{equation}
Here, $d_{vir}=r/R_{vir}$ is the distance from the host-halo centre in
units of  the virial  radius. The parameters  are listed in  Tab.~2 of
\citet{pieri09} for haloes $200$  times denser than the background. In
the  figure, we  plot the  mass-concentration relation  for  clumps at
$10\%$ and $100\%$ of the  virial radius. This proposed relation takes
into  account   that  substructures  with  the  same   mass  are  more
concentrated  if they  are closer  to the  host-halo centre,  which is
necessary for their survival. To  be consistent with our choice of the
virial   overdensity,  we   have  normalized   this   distribution  to
$\Delta_{vir}$. Since for subclumps at the virial radius ($d_{vir}=1$),
the  fit (\ref{pierieq})  follows the  mass-concentration  relation of
isolated  haloes  and  increases  with decreasing  distance  from  the
host-halo centre, we adopt for each model
\begin{equation}
 c_{sub}(m,d_{vir}) = d_{vir}^{-\alpha_R} c_{vir}^I\;, \label{mcsub}
\end{equation}
where  $I=$N,  S,   or  Z,  corresponding  to  C1,   C2  and  C3  (see
Eqs.~\ref{netoeq}, \ref{seljakeq}, \ref{zhaoeq} and \ref{pierieq}).

\subsection{Summary}

Before we  proceed, let us  briefly summarize the main  ingredients of
our extension to the halo model.

\begin{itemize}

\item  The abundance and bias of host haloes  are taken from 
  \citet{sheth99b}.

\item Host haloes and subhaloes are assumed to have NFW density profiles.

\item We study deterministic (D1) and stochastic (D2) models of 
  halo concentration at fixed mass, and explore three different 
  choices for the mean relation between mass and concentration (C1, C2,
  C3). 

\item We study two choices for the density run of subhaloes around the 
  center of the host:  either it is the same NFW profile as the smooth 
  component (S1), or it has the form suggested by \citet{gao04} (S2).  

\item The mass function of subhalos is given by equation~(\ref{subeq}), 
  found in and normalized to numerical simulations.

\item The  concentration of subhaloes increases towards  the centre of
  their   host   halo,  in   agreement   with  numerical   simulations
  \citep{pieri09}.  As we describe below, our halo model includes this
  effect only rather crudely.

\end{itemize}

\section{Non-linear power spectrum}
\label{nonlinearps}

In this Section, we describe the halo-model approach at the non-linear
dark-matter power  spectrum and study  how it changes at  small scales
when different  models for  the mass-concentration relation  are taken
into account. We will also  show how the non-linear power spectrum can
be  decomposed  when  a  realistic  model for  the  substructure  mass
function in host haloes and  two models for their spatial distribution
are  taken   into  account.    The  formalism  for   the  substructure
contribution to the one-halo  term of the non-linear dark-matter power
spectrum has  been developed by \citet{sheth03c}. Here,  we shall also
consider the  clump contributions  to the two-halo  term and  show the
relevant equations.

\subsection{Contributions from host haloes}

In the   halo  model  approach  
\citep{scherrer91,  seljak00, scoccimarro01, cooray02}, 
the two-point dark-matter correlation function is
\begin{equation}
 \xi(\vec{x}-\vec{x}') = \xi_{1H}(\vec{x}-\vec{x}') +
 \xi_{2H}(\vec{x}-\vec{x}')\;,
\end{equation}
where  the first  or Poisson  term describes  the contribution  to the
matter density from individual haloes, while the second term describes
the  contribution from  halo  correlations. While  both terms  require
knowledge  of the  halo mass  function and  their  dark-matter density
profiles, the second needs  also the two-point correlation function of
haloes  of different mass  $\xi_{hh}(r|M_1,M_2)$. Since  the two-point
correlation function on large scales is dominated by the two-halo term
and  has  to  follow  the  linear correlation  function,  a  good  and
convenient way to express the two-halo correlation function is
\begin{equation}
 \xi_{hh}(r|M_1,M_2) \approx b(M_1) b(M_2) \xi_{lin}(r)\,.
\end{equation}
In view  of later  convolutions, it is  convenient to work  in Fourier
space.  We then define the dark matter power spectrum $P(k,z)$ as
\begin{equation}
 P(k,z) = 4 \pi \int \xi(r,z)\frac{\sin(kr)}{kr}r^{2} {\rm d}r \, .
\end{equation}
For further convenience, and as is common in the large scale structure
community, we define the dimensionless quantity
\begin{equation}
 \Delta^{2}(k,z) \equiv \frac{k^{3}P(k,z)}{2 \pi^{2}} \,.
\end{equation}
The dark matter power spectrum can be decomposed as
\begin{equation}
 P(k,z) = P_{1H}(k,z) + P_{2H}(k,z)
\end{equation}
where
\begin{eqnarray}
 P_{1H}(k,z)  &=&   \int  \left(\dfrac{M}{\bar{\rho}}\right)^2  n(M,z)
 \nonumber  \\   &\times&  \int  p(c|M)   u^2(k|c(M))  \mathrm{d}c  \,
 \mathrm{d}M,
\label{pkhalo1}
\end{eqnarray}
and
\begin{eqnarray}
 P_{2H}(k,z)  &=& P_{\rm  lin}(k)  \Big{[} \int  \dfrac{M}{\bar{\rho}}
   n(M,z)\,b(M,z)\nonumber          \\          &\times&          \int
   p(c|M)\,u(k|c,M)\,\frac{b(c,M,z)}{b(M,z)}\,\mathrm{d}c
   \,\mathrm{d}M \Big{]}^2 \label{pkhalo2}
\end{eqnarray}
are the one- and two-halo  contributions. Here, $b(c,M,z)$ is the bias
factor with  respect to the  dark-matter of halos  of mass $M$  at $z$
having concentration $c$.  The dependence  of halo bias on $c$ as well
as $M$ is a simple way  of including the `assembly bias' effect in the
model.   In practice,  we  ignore  this effect,  meaning  that we  set
$b(c,M,z)/b(M,z) = 1$  from now onwards.  Note that the integrals
extend  over  all  collapsed  masses.\footnote{A  lower bound  of zero 
  in the mass integrals actually means that the integration
  starts at $10^{-6}\,M_{\odot}$ \citep{green04,  giocoli08a}.}

The   most   commonly    used   approximations   in   the   literature
\citep{seljak00,cooray02}  use  the  deterministic  model D1  for  the
concentration distribution and consider steep models for the 
mass-concentration relation in order to not loose small scale power.

\begin{figure*}
\centering
\includegraphics[width=13cm]{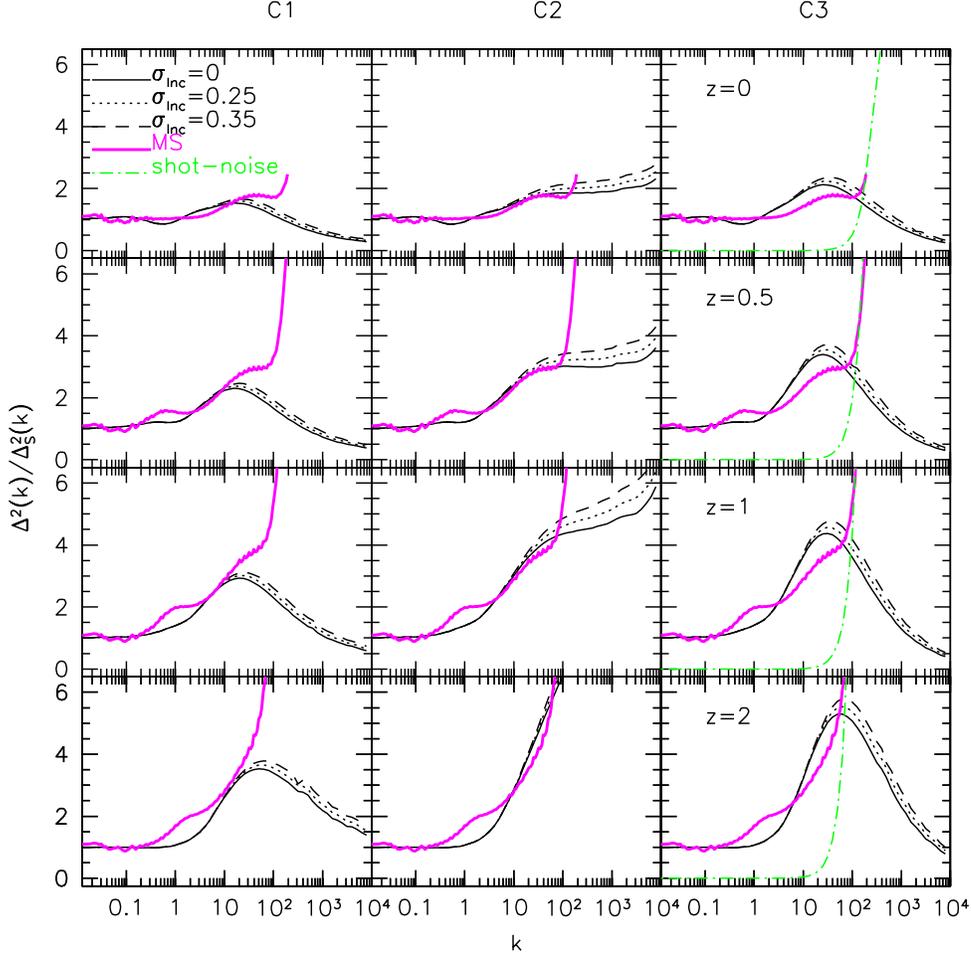}
\caption{\label{PKS}  Ratio of the  non-linear  power  spectrum
  predicted by  the halo  model and the  fit by  \citet{smith03}. From
  left to right,  the panels show results for the  three models of the
  mass-concentration relation,  as described in the text.  From top to
  bottom, we show results at  four different redshifts. In all panels,
  solid  curves show  the  ratio for  the deterministic  concentration
  assignment, while  dotted and dashed curves refer  to the stochastic
  model with widths $\sigma_{\ln  c}=0.25$ and $0.35$. The heavy solid
  curves show the  ratio between the power spectrum  in the Millennium
  Simulation and  \citet{smith03} at the  corresponding redshifts, and
  the dash-dotted line (only in the right panels) shows the shot noise
  in the MS with respect to the fit. \label{RatioSmith}}
\end{figure*}

In Fig.~\ref{PKS}  we show  the non-linear dark-matter  power spectrum
predicted using  the halo  model, as described  above, divided  by the
power-spectrum  fit  by \citet{smith03}  at  four different  redshifts
(from   top   to  bottom)   and   assuming   three   models  for   the
mass-concentration  relation (from  left to  right).  The  solid curve
refers  to   the  analytical   non-linear  power  spectrum   with  the
deterministic concentration  model (D1),  while the dotted  and dashed
curves refer to  the model which adopt the  stochastic model (D2) with
widths $\sigma_{\ln c}=0.25$ and $0.35$, respectively.  The ringing of
the curves for large $k$ is due to the Fourier tranform of the density
profile.  The thick solid curve  refers to the power spectrum measured
in  the  Millennium  Simulation  \citep{springel05b,boylan-kolchin09}.
The dotted-dashed line in the  right panels shows the shot-noise limit
in  the  Millennium  Simulation,  which  we do  not  remove  from  the
numerical power spectrum, at the corresponding redshifts.

From  the Figure,  we notice  that model  C1 reproduces  the  MS power
spectrum  quite well  up to  $k\approx  10$, while  model C2  performs
better.   Model C3 is  also quite  accurate and  does not  decrease at
small  scales, although  it somewhat  overestimates the  power  in the
range  $1   \lesssim  k  \lesssim  50$.   Note   that  the  stochastic
concentration  model (D2)  increases  the power  on  small scales  ($k
\gtrsim 10$) by 15 to 20\% compared to the deterministic (D1) case.

\subsection{Contributions from subhaloes}

\begin{figure*}
\centering \includegraphics[width=13cm]{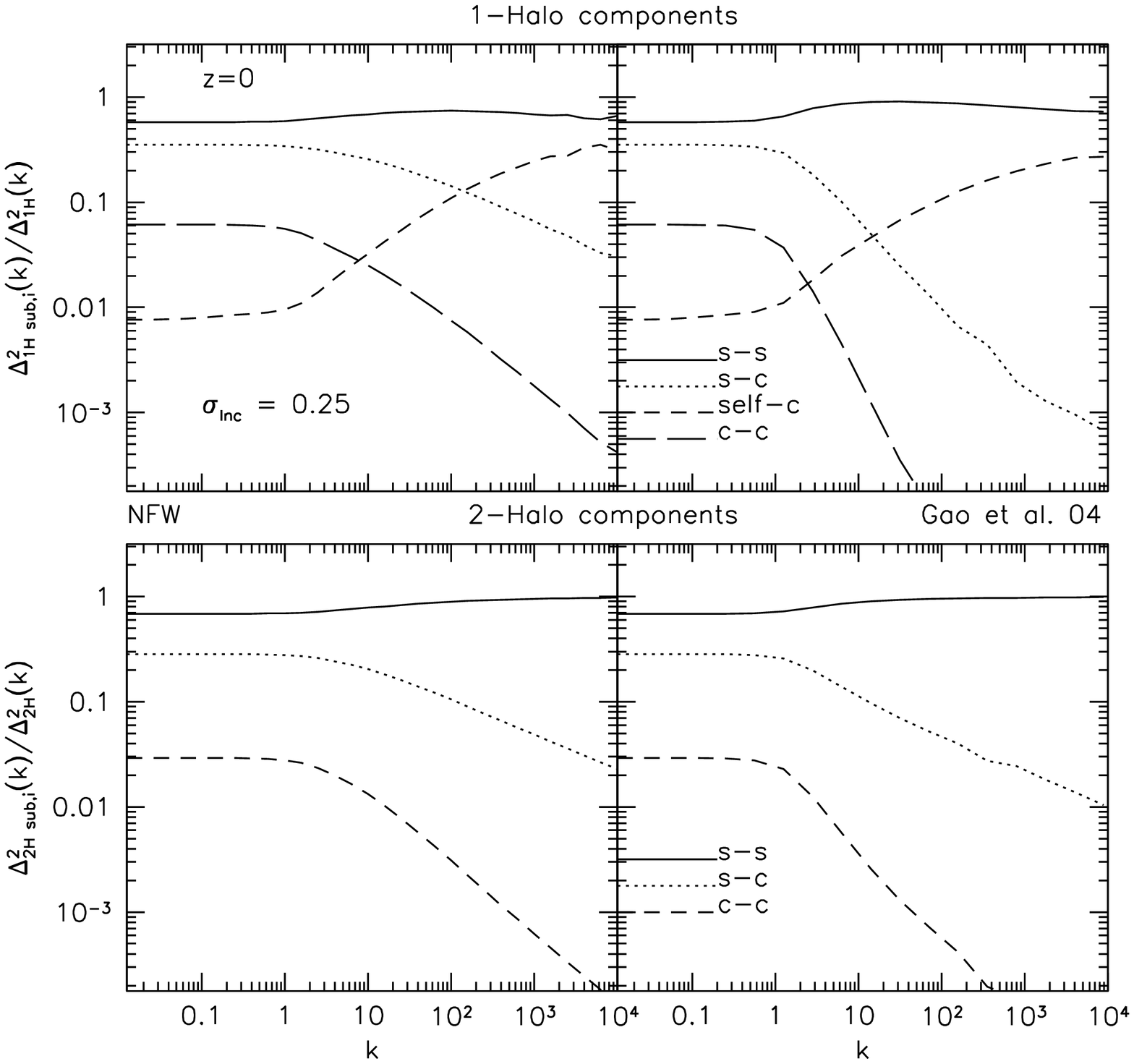}
\caption{Top:   Ratio   of   the   1-halo  components   divided   into
  substructures  (Eqs.~\ref{pksub1},  \ref{pksub2},  \ref{pksub3}  and
  \ref{pksub4})  compared to  the 1-halo  term  (Eq.~\ref{pkhalo1}). A
  width  of $\sigma_{\ln  c}=0.25$  was assumed  in the  concentration
  distribution. In the left and right panels, we adopt the NFW and the
  \citet{gao04} model for the substructure distribution. Bottom: Ratio
  of  the  2-halo   components  (Eqs.~\ref{pksub5},  \ref{pksub6}  and
  \ref{pksub7})       compared      to      the       2-halo      term
  (Eq.~\ref{pkhalo2}).\label{figsubcomppk}}
\end{figure*}

Considering the  contribution from substructures, the  1-halo term can
be  further  decomposed  into   four  contributions  from  the  mutual
correlations between smooth and  clump components as explained in what
follows.

\begin{itemize}
\item [($1$)] Smooth-smooth correlation:
\begin{eqnarray}
 P_{1H,ss}(k,z) &=&  \int \left(\dfrac{M}{\bar{\rho}} \right)^2 n(M,z) 
\\ &\times& \nonumber
\int \left(\dfrac{M_{sm}}{M}\right)^2 u^2(k|c(M_{sm})) p(c|M) \mathrm{d}c \, \mathrm{d}M
\label{pksub1}
\end{eqnarray}

\item[($2$)] Smooth-clump correlation:
\begin{eqnarray}
 P_{1H,sc}(k,z) &=& 2 \int \left( \dfrac{M}{\bar{\rho}}\right)^2 n(M,z) 
 \int \dfrac{M_{sm}}{M} u(k|c(M_{sm})) \, \nonumber \\ &\times& U_s(k|c(M)) 
 \mathcal{I}_{c(M)}(k,z) p(c|M) \mathrm{d}c \, \mathrm{d}M 
\label{pksub2}
\end{eqnarray}

\item[($3$)]  Two-point correlation  between different  clumps  in the
  same halo:
\begin{eqnarray}
 P_{1H,cc}(k,z) &=&  \int \left( \dfrac{M}{\bar{\rho}} \right)^2 n(M,z) \\ \nonumber
 &\times& \int U^2_s(k|c(M)) \, \mathcal{I}^2_{c(M)}(k,z) p(c|M)
 \mathrm{d}c \, \mathrm{d}M
\label{pksub3}
\end{eqnarray}

\item[($4$)] Two-point correlation between pairs in the same clump:
\begin{eqnarray}
 P_{1H,self-c}(k,z) &=&  \int \left( \dfrac{M}{\bar{\rho}} \right)^2
 n(M,z) \\ 
&\times&
\int
 \mathcal{J}_{c(M)}(k,z) p(c|M) \mathrm{d}c\, \mathrm{d}M \nonumber
\label{pksub4}
\end{eqnarray}
\end{itemize}

Here, we define the smooth mass as
\begin{equation}
 M_{sm} = (1-f_s) \, M ,
\end{equation}
which depends on the host halo concentration (see eq.~\ref{subeq}),
while $\mathcal{I}$ and $\mathcal{J}$ are the functions
\begin{equation}
 \mathcal{I}_{c(M)}(k,z) = \int_0^M \dfrac{m}{M} \int u_m(k,c)
 \frac{\mathrm{d}N(c(M),z)}{\mathrm{d}m} p(c|M)
 \mathrm{d}c\,\mathrm{d}m
\label{intI}
\end{equation}
and
\begin{equation}
 \mathcal{J}_{c(M)}(k,z) = \int_0^M \left(\dfrac{m}{M}\right)^2 \int u^2_m(k,c)
 \frac{\mathrm{d}N(c(M),z)}{\mathrm{d}m} p(c|M)\mathrm{d}c\,
 \mathrm{d}m\,,
\end{equation} 
where we  also take into account the  scatter in at fixed  mass in the
subhalo  mass-concentration  relation  when  the stochastic  model  is
considered.  The subhalo concentration depends on its mass and also on
its radial  distance from the  host halo centre  (see eq.\ref{mcsub}).
The  mass-concentration relation  adopted  is consistent  with the  one
assumed   for   isolated   host   haloes,  such   that   the   subhalo
mass-concentration  relation,  for  distance  larger then  the  virial
radius of  the host follows the halo  mass-concentration relation.  To
account  for  the  radial  distance  dependance we  randomly  sample  the
substructure  density profile  distribution in  the  host, consistently
with the model adopted (NFW  or \citet{gao04}), and locate the subhalo
at a distance  $d_{vir}$ from the host centre  ($d_{vir}$ is expressed
in unit of the host halo virial radius).  Note that we do not consider
scatter in $f_s$ at fixed $c$ and $M$.

\begin{figure*}
\centering
\includegraphics[width=13cm]{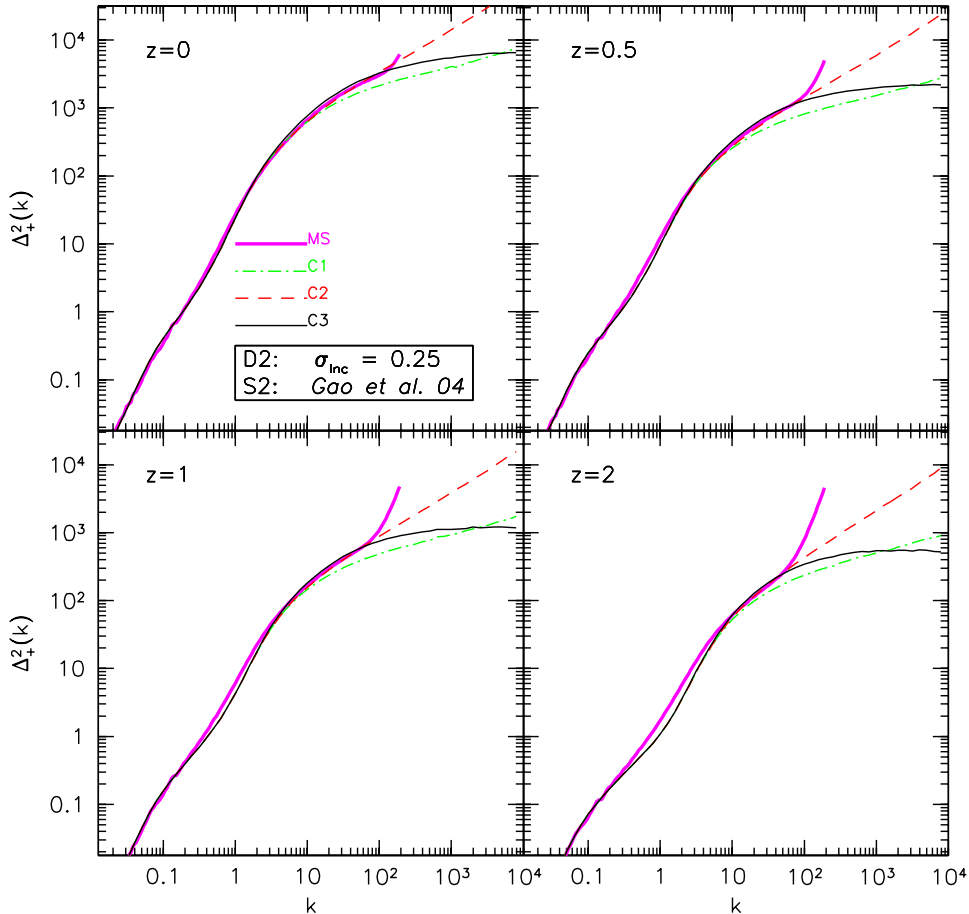}
\caption{Non-linear dark-matter power spectrum, with substructures, at
  four different redshifts. We show three models referring to different
  mass-concentration relations.  The heavy  solid line shows  the data
  from  the  Millennium Simulation.  The  width  of the  concentration
  distribution  is  assumed  to  be  $\sigma_{\ln  c}=0.25$,  and  the
  substructure distribution in the  host halo is modeled according to
  \citet{gao04}.  For all models,  the mass  distribution in  both the
  host   haloes   and  in   their   substructures   follows  the   NFW
  profile.\label{fZhao4z}}
\end{figure*}

Likewise,  the large-scale  (2-Halo)  term can  be further  decomposed
according  to  the different  correlations  between  smooth and  clump
components. In this case, we have the three contributions

\begin{itemize}
\item[($1$)] Smooth-smooth component correlation on large scales:
\begin{equation}
 P_{2H,ss}(k,z) = P_{lin}(k) {\mathcal{S}^I}^2(k,z)
\label{pksub5}
\end{equation}

\item[($2$)] Smooth-clump correlation:
\begin{equation}
P_{2H,sc}(k,z) = 2 P_{lin}(k) \mathcal{S}^I(k,z) 
\mathcal{C}^I(k,z)\,
\label{pksub6}
\end{equation}

\item[($3$)] Clump correlation:
\begin{equation}
P_{2H,cc}(k,z) = P_{lin}(k) {\mathcal{C}^I}^2(k,z)
\label{pksub7}
\end{equation}
\end{itemize}

Here,
\begin{eqnarray}
 \mathcal{S}^I(k,z) &=& \int \dfrac{M}{\bar{\rho}} n(M,z) b(M,z) \nonumber \\ 
 &\times& \int \dfrac{M_{sm}}{M} u(k,c(M_{sm})) p(c|M) \mathrm{d}c \, \mathrm{d}M,
\label{sieq}
\end{eqnarray}
and
\begin{eqnarray}
 \mathcal{C}^I(k,z) &=& \int \dfrac{M}{\bar{\rho}} n(M,z) b(M,z) \nonumber \\ &\times& \int
 \mathcal{I}_{c(M)}(k,z) U_s(k|c(M)) p(c|M) \mathrm{d}c \,
 \mathrm{d}M. \label{cieq}
\end{eqnarray}
Recall that  $I = $  N,S, or  Z, corresponding to  C1, C2 and  C3 (see
Eqs.~\ref{netoeq}, \ref{seljakeq}, \ref{zhaoeq} and \ref{pierieq}).

The  full non-linear  dark-matter power  spectrum  is the  sum of  all
components,
\begin{equation}
 P_+(k,z) = \sum_{i=1}^4 P_{1H,i}(k,z) + \sum_{i=1}^3 P_{2H,i}(k,z)\,.
\end{equation}

In  Fig.~\ref{figsubcomppk}  we  show  the  ratio  of  the  individual
contributions  to the  1- and  2-halo  components.  We  set the  width
$\sigma_{\ln c}=0.25$ for the concentration distribution, and adopt C3
for the mass-concentration relations.   All functions depending on the
concentration are estimated by  marginalizing over $c$.  Models S1 and
S2 for  the subclump distribution within  the host are  adopted in the
different  panels.  Note that  in both  the 1-  and 2-halo  terms, the
dominant contribution is due to pairs in which both members are in the
smooth component; smooth-clump  and clump-clump pairs never contribute
more than  10\% of the signal,  and they become even  less dominant at
large $k$.   The transition  scale is around  $k \approx 1$,  since it
represents the  size of a typical  collapsed halo in  which the clumps
are located. For $k \gtrsim  1$ the smooth-clump and clump-clump terms
start  to  drop  down  due  to the  density  profile  distribution  of
subhaloes in the host system.   In other words, for scale smaller then
1 Mpc  $h^{-1}$ the two-point probability  has a threshold  due to the
typical distribution scale  of the clumps within the  virial radius of
the  host\footnote{We recall that  the power  spectrum is  the Fourier
  tranform of  the two-point correlation  function.}.  Comparing right
and left panels we notice that the power-spectrum components are quite
sensitive to the more rapid decrease of the Fourier transform model S2
(right panels) compared to model  S1 (left panels), both in the 1-Halo
and the 2-Halo terms.  However, the contribution from pairs within the
same subclump (self-clump) increases with $k$ reaching values of order
$50\%$ at $k\approx  10^4$, this is due to  the matter density profile
distribution in small clumps, more and more concentrated as their mass
decrises. The probability  to find a pair of points  of the same clump
becames larger and larger as the scale decreases -- so for large value
of $k$.

In Fig.~\ref{fZhao4z},  we show the full non-linear  power spectrum at
four different redshifts. We include  results for models C1, C2 and C3
and  for $\sigma_{\ln  c}=0.25$.  We  adopt  the NFW  profile for  the
matter-density  distribution  in  both   the  host  haloes  and  their
substructures, and model S2 for the radial subclump distribution within
the host haloes.

\begin{figure}
\centering
\includegraphics[width=\hsize]{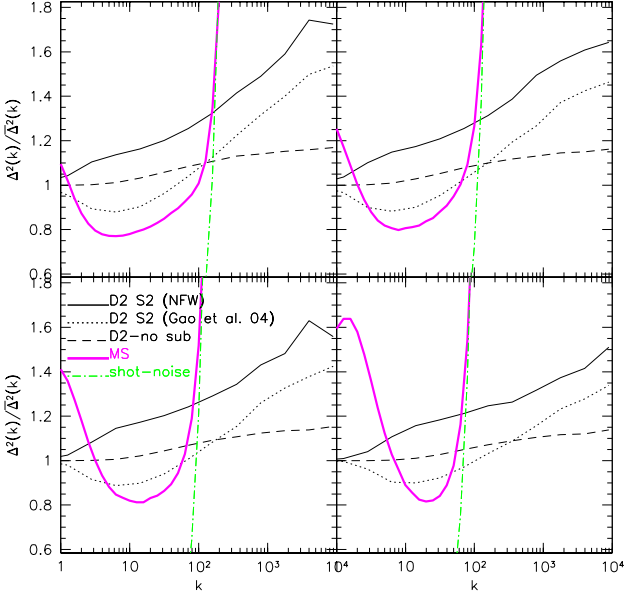}
\caption{Ratio between  the predicted power-spectrum  in our extension
  of the  halo model, which includes substructures  and the stochastic
  assignment  of halo  concentrations, to  the power  spectrum  in the
  standard  implementation  of  the  halo model,  which  ignores  halo
  substructures and assigns concentrations deterministically.  For the
  log-normal scatter in  concentration we adopt $\sigma_{\ln c}=0.25$.
  The solid and  dotted curves assume an NFW  or a \citet{gao04} model
  for  the  subclump distribution.   The  dashed  line shows the 
  predicted ratio when we allow scatter  in the  concentrations 
  but assume smooth halos with no substructure.  The  thick
  magenta  line  shows the  power-spectrum  ratio  for the  Millennium
  Simulation and the dash-dotted line shows the shot-noise.\label{ratioSZ}}
\end{figure}

In Fig.~\ref{ratioSZ}  we show the  ratio between the  full non-linear
power  spectrum  with  $\sigma_{\ln  c}=0.25$, considering  models  S1
(solid) and S2  (dotted), and the standard halo model  -- 1 and 2-Halo
terms and no scatter in  concentration, at four different redshifts as
in  Figure \ref{fZhao4z}.   This figure  quantifies the  importance of
substructures  and  the  concentration  scatter  in  the  halo  model.
Including   both   effects   increases   the  small-scale   power   by
$\approx30\%$.  On  the large scales where the  2-Halo term dominates,
accounting  for  or  ignoring  substructure makes  no  difference,  as
expected  \citep{sheth03c}.  In  the  figure we  also  show the  ratio
between  the  power spectra  in  the MS  (thick  solid  line) and  the
standard halo  model with stochastic  concentration assignment (dashed
line),  compared  to the  standard  halo  model without  concentration
scatter. The dash-dotted line show the shot-noise.

\section{Cross correlations}
\label{crosscorrelations}

We now  describe halo models  of the cross correlation  between haloes
and mass,  as well as between  substructures and mass.   We shall show
that while  the cross correlation with the  smooth component dominates
the signal  on large scales,  the self-cross correlation  dominates on
small scales.  We shall  also quantify the  scales on  which different
host-halo  masses  and  substructures  contribute  significantly.   In
contrast to  \citet{hayashi08}, who presented  a simple model  for the
auto- and  cross-correlation functions, we study  all contributions to
the non-linear  power spectrum explicitly.  As  for the power-spectrum
in  the  previous  section,  our  model of  the  1-halo  term  follows
\citet{sheth03c}.

\begin{figure*}
\centering
\includegraphics[width=13cm]{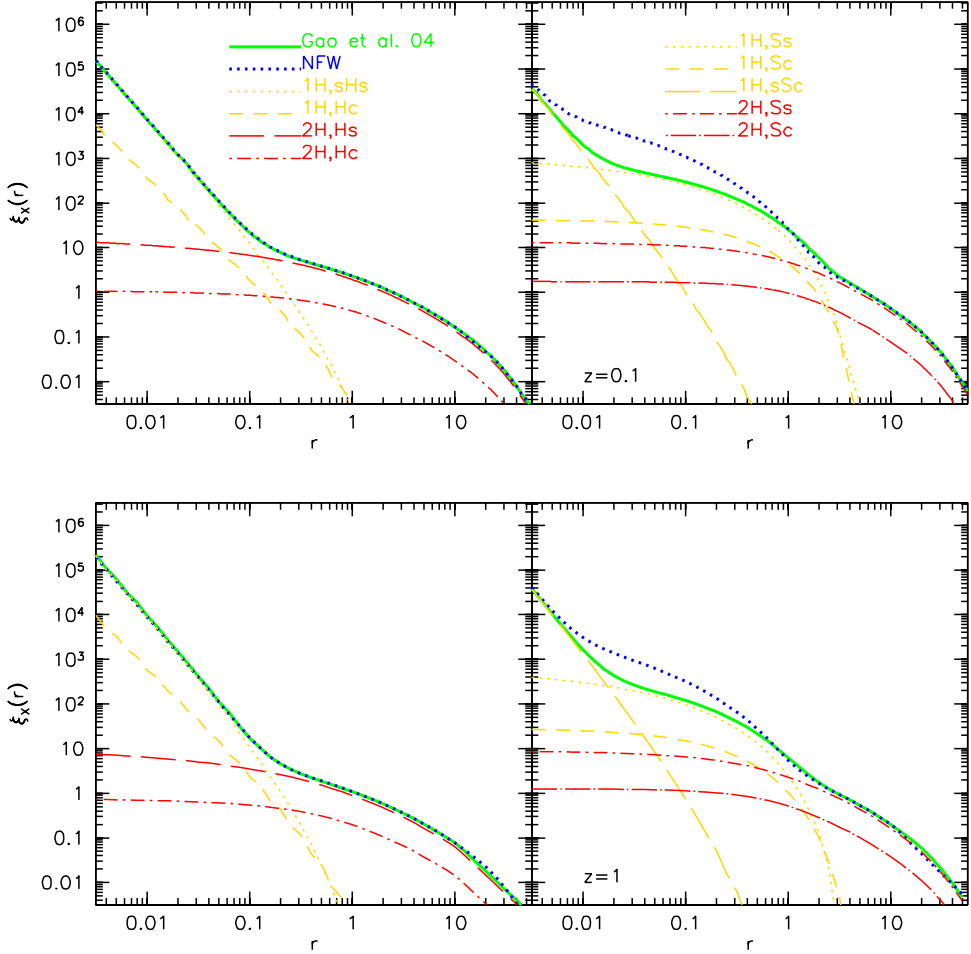}
\caption{Halo and subhalo-mass  cross-correlations at redshift $z=0.1$
  (top) and $z=1$  (bottom). In each panel, we  show the contributions
  of haloes and substructures to the 1- and 2-halo terms. The left and
  right panels show the  cross-correlation between haloes and mass and
  between  substructures   and  mass,  respectively.    We  adopt  the
  \citet{gao04} model for the  radial subclump distribution in the host
  halo,  \citet{zhao09}   for  the  mass-concentration   relation  and
  $\sigma_{\ln c}=0.25$ for its width. The heavy solid curve shows the
  sum of all contributions. For comparison, we also show in each panel
  the total  cross-correlation function when  the spatial distribution
  of  subclumps  in  the  host  follows  the  NFW  model  (heavy-dotted
  curve).\label{crossfig1}}
\end{figure*}

\subsection{Cross-correlation of haloes and mass}

We begin by  writing down the contributions to the  1-halo term of the
cross-correlation power spectrum between haloes and matter. Sitting at
the center  of a host  halo, and assuming  that the matter  is divided
into  smooth and clumpy  components, the  halo-smooth self-correlation
and the halo-clump cross-correlation can be written as
\begin{eqnarray}
 P^{\times}_{1H,sHs}(k,z) &=&  \int \dfrac{M}{\bar{\rho}} n(M,z)
 \nonumber \\ &\times& \int \dfrac{M_{sm}}{M} u(k|c(M_{sm})) p(c|M) \mathrm{d}c \,\mathrm{d}M,
\label{crossh1}
\end{eqnarray}
and
\begin{eqnarray}
 P^{\times}_{1H,Hc}(k,z) &=& \int \dfrac{M}{\bar{\rho}} n(M,z)
 \nonumber \\ &\times& \int U_s(k,c) \mathcal{I}_{c(M)}(k,z) p(c|M)
 \mathrm{d}c\,\mathrm{d}M.
\label{crossh2}
\end{eqnarray}
We  recall  that  $M_{sm}$  is  the  smooth mass  of  the  host  halo,
$u(k|c(M_{sm}))$ and  $U_s(k|c(M))$ are the Fourier  transforms of the
density profiles of the smooth component and of the clump distribution
respectively. Moreover, $n(M,z)$ is the halo mass function at redshift
$z$ and $\mathcal{I}_{c(M)}$ is given by Eq.~(\ref{intI}).

Regarding  the contributions  to large  scales (the  2-halo  term), we
start from the  same idea, i.e.~we imagine sitting at  the center of a
host  halo  and  cross-correlate   with  the  smooth  and  the  clumpy
components of  a \emph{distant}  halo. However, in  this case  we also
have to  take the halo bias  relative to the  dark matter distribution
into   account,  as   well   as  the   non-linear  dark-matter   power
spectrum. The smooth and clump cross-correlation power spectra are
\begin{equation}
 P^{\times}_{2H,Hs}(k,z) = \frac{P_{lin}(k,z)}{\bar{n}_h}
 \int n(M,z) b(M,z) \mathrm{d}M \mathcal{S}^I(k,z)
\label{crossh3}
\end{equation}
and
\begin{equation}
P^{\times}_{2H,Hc}(k,z) = \frac{P_{lin}(k,z)}{\bar{n}_h}
\int n(M,z) b(M,z) \mathrm{d}M \mathcal{C}^I(k,z),
\label{crossh4}
\end{equation}
with $\mathcal{S}^I$ and $\mathcal{C}^I$ given by Eqs.~(\ref{sieq}) and
(\ref{cieq}). 

\subsection{Cross-correlation of subhaloes and mass}

Let us now place ourselves at  the center of a clump and determine the
cross-correlation on small scales. In  this case, the 1-halo term will
be  the sum  of three  components: (i)  the self-correlation  with the
substructure  mass; (ii) the  cross-correlation with  the mass  in the
other substructures  contained in  the same host  halo, and  (iii) the
cross-correlation with the smooth component. These three terms are
\begin{eqnarray}
 P^{\times}_{1H,Ss}(k,z) &=& \frac{1}{\bar{n}_s} \int \dfrac{M}{\bar{\rho}}
 n(M,z) \int \dfrac{M_{sm}}{M} u(k|c(M_{sm})) \nonumber \\ &\times& U_s(k|c(M)) 
 N_s(c(M),z) p(c|M) \mathrm{d}c \, \mathrm{d}M,
\label{crossc1}
\end{eqnarray}
further
\begin{eqnarray}
 P^{\times}_{1H,Sc}(k,z) &=& \frac{1}{\bar{n}_s} \int \dfrac{M}{\bar{\rho}}
 n(M,z) \int U^2_s(k|c(M)) \nonumber \\ &\times&
 \mathcal{I}_{c(M)}(k,z) N_s(c(M),z) p(c|M) \mathrm{d}c \,
 \mathrm{d}M
\label{crossc2}
\end{eqnarray}
and 
\begin{equation}
 P^{\times}_{1H,sSc}(k,z) = \int \dfrac{M}{\bar{\rho}}
n(M,z) \int \mathcal{I}_{c(M)}(k,z) p(c|M) \mathrm{d}c \, \mathrm{d}M,
\label{crossc3}
\end{equation}
where we have underlined that  $N_s$, the number of substructures in a
$M$-halo,  depends  on  the  mass  through the  concentration  and  on
redshift.  For the  2-halo term,  we have  two equations  analogous to
those of  the halo-matter cross-correlation, and we  need to integrate
over the substructure mass function. We find
\begin{eqnarray}
 P^{\times}_{2H,Ss}(k,z) &=& \frac{P_{lin}(k,z)}{\bar{n}_s}
 \int n(M,z) b(M,z) \int N_s(c(M),z) \nonumber \\ &\times& U_s(k|c(M))
 p(c|M) \mathrm{d}c\,\mathrm{d}M \, \mathcal{S}^I(k,z),
\label{crossc4}
\end{eqnarray}
and
\begin{eqnarray}
 P^{\times}_{2H,Sc}(k,z) &=& \frac{P_{lin}(k,z)}{\bar{n}_s}
 \int n(M,z) b(M,z) \int N_s(c(M),z) \nonumber \\ &\times& U_s(k|c(M))
 p(c|M) \mathrm{d}c\, \mathrm{d}M \, \mathcal{C}^I(k,z).
\label{crossc5}
\end{eqnarray}
We  show  the  results  in  real  space  for  each  component  of  the
cross-correlation after the inverse Fourier transform
\begin{equation}
 \xi_{i,\times}(r,z) = \frac{1}{2 \pi^2} \int P_{iH,k}(k,z) \frac{\sin (k r)}{k r}
 k^2 \mathrm{d}k\;,
\end{equation}
where   the  index   $i$   runs  over   all   halo  and   substructure
cross-correlation terms.   Going from Fourier  to real space,  some of
the  1-halo   terms  are  simple,  because  we   are  inverse  Fourier
transforming the NFW density profile  and the density run of subclumps
in the  host (see eq.~\ref{eqclumpdist}).  For the model S1,  in which
subclumps follow  the dark matter distribution, terms  like $uU_s$ are
the  convolution of  two NFW  profiles, so  they can  also  be written
analytically \citep{sheth01c}.

\begin{figure*}
\centering
\includegraphics[width=13cm]{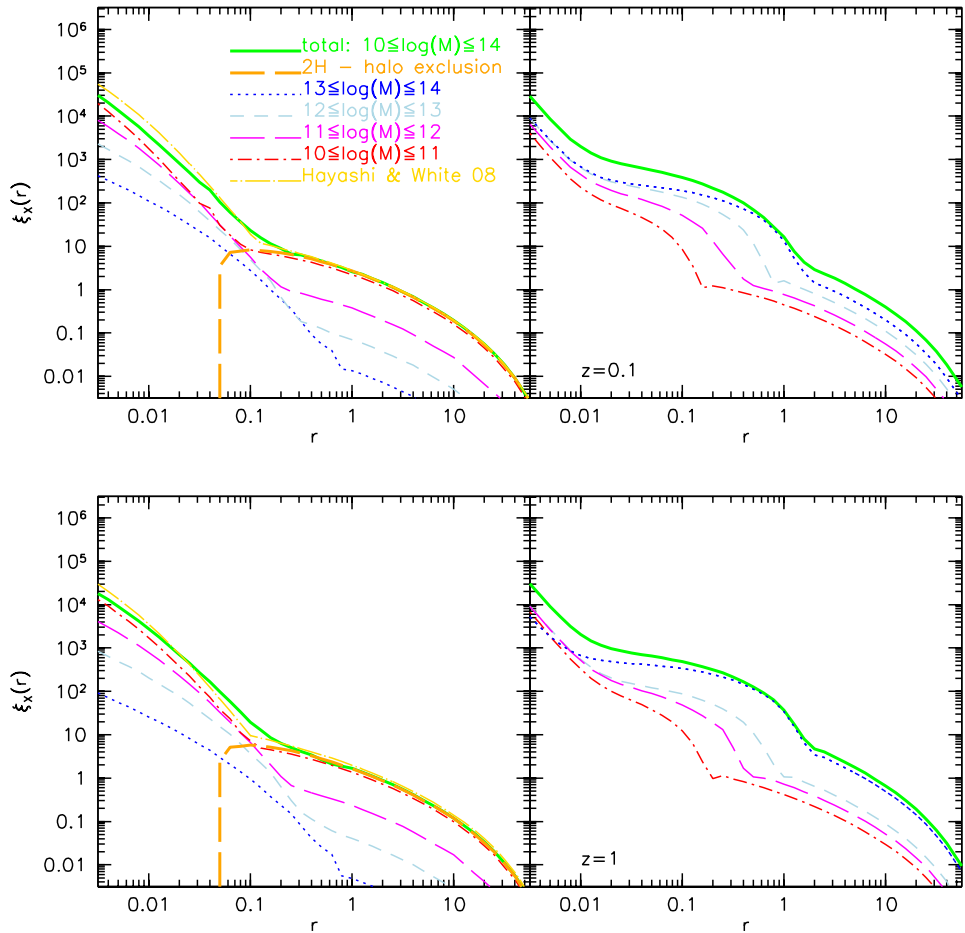}
\caption{Halo and subhalo-mass cross-correlations at $z=0.1$ and $z=1$
  separating  the contributions from  different host-halo  masses. The
  left panels show the  halo-mass cross-correlation.  The 2-halo terms
  in different mass  bins have been normalized to  the total number of
  haloes  between  $10^{10}$  and $10^{14}\,M_{\odot}/h$.   The  right
  panels show the subhalo-mass cross-correlation considering clumps in
  different  host-halo  masses. The  curves  in  this  case have  been
  normalized to  the mean  number of clumps  in haloes with  mass from
  $10^{10}$ and $10^{14}\,M_{\odot}/h$.\label{crossfig2}}
\end{figure*}

In Fig.~\ref{crossfig1}, we  show the cross-correlation between haloes
and matter (left  panels) as well as between  substructures and matter
(right panels) at the two different redshifts $z=0.1$ and $z=1$. These
represent   the   mean   redshifts    of   galaxies   in   the   SDSS
\citep{adelman-mccarthy06,   wang07b}  and  in   typical  weak-lensing
surveys,  respectively. In the  left panels,  we show  the halo-matter
cross-correlation function considering the  two components both in the
1- and 2-halo terms. We  adopt models C3, D2 with $\sigma_{ln c}=0.25$
and S2. The solid and  dotted curves show the total cross-correlation,
with distributions  S1 and  S2 for placing  substructures in  the host
halo.   Notice that, in  the host  mass cross-correlation,  the latter
choice  does  not  influence  the  total.   On  the  other  hand,  the
cross-correlation between  substructures and mass is  sensitive to the
choice of  the radial subclump  distribution.  In the right  panels, we
show the cross correlation between clump and mass.  Here, too, we show
the three components of the 1-halo  term and the two components of the
2-halo term.  The solid and dotted curves are, respectively, the total
predicted signal  using models S1 and  S2.  From the  figure we notice
that substructures contribute a signature  on scales of order $30$ kpc
or so.

In   the   left   panels   of  Fig.~\ref{crossfig2},   we   show   the
cross-correlation  function between haloes  and matter,  splitting the
contributions  into different  host-halo mass  bins. The  integrals in
Eqs.~(\ref{crossh1}, \ref{crossh2}, \ref{crossh3}) and (\ref{crossh4})
have been  performed between the lower  and upper bounds  of each bin,
while   $\mathcal{S}^I(k,z)$   and   $\mathcal{C}^I(k,z)$  have   been
integrated  over all  masses. The  solid  line shows  the total  power
spectrum from haloes in  the entire mass range, i.e.~between $10^{10}$
and  $10^{14}\,M_{\odot}/h$. The  curves  are normalized  to the  mean
number of haloes in the entire range.

The long-dashed-dotted  curve in the  left panels, shows the  model of
\citet{hayashi08}. In their  model the 1-halo term is  due to the dark
matter density  profile of the host  halo, while the two  halo term by
the product between bias  factor and the mass autocorrelation function
predicted by  the linear theory. In  their model we  consider the mean
mass predicted by the \citet{sheth99b} mass function between $10^{10}$
and $10^{14}\,M_{\odot}/h$.

In the  right panels, we show  the substructure-mass cross-correlation
considering clumps in the same host-halo mass bins.  These curves have
been  normalized  by the  mean  number  of  clumps in  haloes  between
$10^{10}$  and  $10^{14}\,M_{\odot}/h$.   In  this case,  showing  the
predicted contribution  from different mass bins, we  account for halo
exclusion in the halo-mass cross correlation. An accurate treatment of
halo  exclusion  may  be   numerically  tedious.  Here,  we  adopt  an
approximate  approach in  which  the 2-halo  term  vanishes on  scales
smaller than the virial radius of  the mean mass in the bin considered
(see e.g. \citep{cacciato09} for a similar approach in the galaxy-dark
matter cross-correlation). Thus, if $M_1$  and $M_2$ are the lower and
upper bounds of the bin, we can write
\begin{equation}
  1+ \xi^{h-excl}_{2H}(r|M_1,M_2) \approx \left[ 1 + \xi_{2H}(r|M_1,M_2)\right] \times W(r|R_{vir})\,
\end{equation}
where $R_{vir}$  is the virial radius  of the mean  mass between $M_1$
and  $M_2$ which is used as a typical cut-off  scale of  the 2-halo
cross-correlation  signal here. $W(r|R_{vir})$  is  a window  function
equal to $1$ if $r>R_{vir}$ and $0$ for $r\leq R_{vir}$.

The curves in the left panels of Fig.~\ref{crossfig2} account for this
effect.   The signal  near the  transition between  the 1-  and 2-halo
terms is  reduced by $30\%$ due  to halo exclusion.   The heavy dashed
line   shows   the   total   2-halo   term   between   $10^{10}$   and
$10^{14}\,M_{\odot}/h$, taking halo exclusion into account.  The curve
tends  to zero  at a  scale corresponding  to the  virial radius  of a
$10^{10}\,M_{\odot}/h$  halo.   The 2-halo  term  integrals have  been
divided into  different bins,  for each of  which we assign  a cut-off
scale corresponding to the virial  radius of a mean halo.  To estimate
the    total    signal    from    haloes   between    $10^{10}$    and
$10^{14}\,M_{\odot}/h$  we sum the  contributions from  the individual
mass bins.   In the clump-mass  cross-correlation, we ignore  the halo
exclusion  since  the  transition   between  the  1-  and  the  2-halo
contributions occurs  at larger scales,  hence the exclusion  will not
change the signal at all.

\section{Summary and Conclusions}
\label{sandc}

We have shown how the  halo-model formalism can be extended to account
for scatter in halo concentration  at fixed mass, and for the presence
of  substructures   in  dark   matter  haloes.   Differences   in  the
mass-concentration relation  do affect the  predicted non-linear power
spectrum.   We quantified  this by  considering three  models  for the
$M$-$c$ relation, as  well as a deterministic and  a stochastic models
for  the concentration  at  fixed mass.   Accounting for  substructure
means that  the 1-halo and 2-halo  terms should be written  as sums of
four and  three types of  pairs, respectively.  If one  uses realistic
models for these different pair-types, then the halo model calculation
is in reasonable agreement with the non-linear power spectrum measured
in the Millennium Simulation, over a range of redshifts.

The  cross-correlation between halos  and mass  can also  be estimated
using the halo-model formalism. We  have shown how, also in this case,
substructures can be taken  into account and how the cross-correlation
signal  between   clumps  and  mass   can  be  split   into  different
contributions.

The agreement  with simulations  is not yet  at the percent  level, so
there  is room  for improvement.   The simulations  show an  excess at
around $k\sim 1$ relative to our model -- this is almost certainly not
due problems with our substructure  calculation, since it is on larger
scales than those on which substructures dominate.  The discrepancy is
also  unlikely to be  due to  our decision  to ignore  `assembly bias'
effects.  Rather it may be due to the fact that our 2-halo term is too
crude  -- on these  smaller scales,  one must  include the  effects of
nonlinear bias, perhaps following \citet{smith07}.

\begin{itemize}
\item Our formulation of the halo model assumes that the mass fraction
  in   substructures  is   a  deterministic   function  of   the  halo
  concentration: in practice, there  is a distribution of $f_s$ values
  at fixed $c$  and $M$.  Allowing for this  stochasticity simply adds
  one more integral in most of our expressions, that can matter at the
  percent level on the small  scales of interest here (for essentially
  the same reason that scatter in $c$ and fixed $M$ matters at the ten
  percent level on small scales).
\item We only include some of the effects of the known correlation 
between subhalo concentration and distance from the center of the host.  
Specifically, our current implementation accounts for the fact that 
the spatial dependence means subhalo concentrations (at fixed mass) 
are stochastic, but it does not include the fact that this is 
correlated with distance from host halo center.  Implementing a 
spatially dependent concentration is difficult in Fourier space, so 
doing so in configuration is probably the best way forward.  
\item We also do not account for the fact that subhalos themselves 
have substructure.  This will matter most for the contribution which 
comes from pairs which are in the same subhalo (for the same reason 
that substructure is not important for the 2-halo term).  
\item Halo exclusion matters for the cross-correlation signal.  
There are exclusion effects associated with the subhalos too, which 
we currently ignore.  
\item Although the sum of the smooth and clumpy components should 
give an NFW profile, our current implementation assumes the smooth 
component follows and NFW profile, whereas the clumpy one does not.  
Hence there is no guarantee that the sum of the two actually is 
NFW at high precision.  While this is relatively easy to 
fix (simply define the profile of the smooth component to be the 
required NFW minus the clumpy component), we have not done so.  
\item And finally, at this level of precision, halo shapes also matter.  
In particular, the distribution of $f_s$ and $c$ at fixed $M$ may 
depend on halo shape -- once this has been quantified in simulations, 
it may become necessary to extend our formalism to include shapes.  
\end{itemize}

The  formalism presented  in  this  paper can  be  used in  subsequent
studies  to  make more  realistic  predictions  for the  galaxy-galaxy
lensing signals  and for the  lensing convergence power  spectrum.  By
identifying the  different contributions due to the  smooth and clumpy
components  of the  dark-matter haloes,  this  will also  allow us  to
quantify the contributions the  convergence power spectrum not just by
different halo masses, but also by different substructures.

\section*{Acknowledgements}
GC thanks Francesco Pace, Giuseppe    Tormen and Andrea Macci\`o   for
useful discussions.  We thank  also Donghai Zhao,  Ana Valente and the
anonymous Referee who  helped us to  improve the  final presentaion of
this  paper.  Thanks  to Mike  Boylan-Kolchin for  providing us the MS
power spectrum.   GC and MB are  supported by EU-RTN ``DUEL''.  RKS is
supported in  part by NSF  AST-0908241.  MC  acknowledges support from
the German-Israeli Foundation (GIF) I-895-207.7/2005

\bibliographystyle{mn2e}

\label{lastpage}
\end{document}